\documentclass[12pt]{article}
\usepackage{amsmath}
\usepackage{amsfonts}
\usepackage{amssymb}
\usepackage{latexsym}

\usepackage[dvips]{graphicx}
\usepackage{epsf}

\allowdisplaybreaks \numberwithin{equation}{section}

\let\a=\alpha \let\b=\beta         \let\d=\delta
       \let\th=\theta     
 \let\l=\lambda  \let\m=\mu        \let\n=\nu   
    \let\r=\rho     
\let\s=\sigma         
          
\let\G=\Gamma        \let\L=\Lambda 
                
        \let\eps=\epsilon

\newcommand{\cA}{\mathcal{A}}  \newcommand{\cB}{\mathcal{B}}  
    
\newcommand{\cJ}{\mathcal{J}}  \newcommand{\cL}{\mathcal{L}}  
    \newcommand{\cP}{\mathcal{P}}
\newcommand{\cQ}{\mathcal{Q}}  \newcommand{\cR}{\mathcal{R}}  
  \newcommand{\cU}{\mathcal{U}}  

  \newcommand{\Rom}{\mathbb{R}}
  
\newcommand{\tr}{{\mathrm{tr}}} \newcommand{\Tr}{{\mathrm{Tr}}}  
    \newcommand{\mY}{\mathrm{Y}}

\newcommand{\gb}{\bar{g}}  \newcommand{\hb}{\bar{h}}  \newcommand{\mb}{\bar{m}}

\newcommand{\Cb}{\bar{C}}  \newcommand{\Db}{\bar{D}}  
\newcommand{\Gb}{\bar{\Gamma}}    \newcommand{\Lb}{\bar{L}}
\newcommand{\Gbb}{\bar{G}} 
  \newcommand{\Rb}{\bar{R}}

  \newcommand{\tG}{{\widetilde{\Gamma}}}

\newcommand{\Rh}{\widehat{R}}

\newcommand{\half}{\tfrac{1}{2}}		\newcommand{\p}{\partial}
\newcommand{\bBox}{\overline{\square}}

\newcommand{\Sg}{\sqrt{g}}
\newcommand{\Sgb}{\sqrt{\bar{g}}}

\let\=\equiv
\let\io=\infty

\newcommand{\be}{\begin{equation}}
\newcommand{\ee}{\end{equation}}

\newcommand{\bea}{\begin{align}}
\newcommand{\eea}{\end{align}}

\begin{document}
\thispagestyle{empty}
\begin{flushright} \small
MZ-TH/10-08
\end{flushright}
\bigskip

\begin{center}
 {\LARGE\bfseries  Matter Induced Bimetric Actions \\[1.2ex]
for Gravity  
}
\\[10mm]
Elisa Manrique, Martin Reuter and Frank Saueressig \\[3mm]
{\small\slshape
Institute of Physics, University of Mainz\\
Staudingerweg 7, D-55099 Mainz, Germany \\[1.1ex]
{\upshape\ttfamily manrique@thep.physik.uni-mainz.de} \\
{\upshape\ttfamily reuter@thep.physik.uni-mainz.de} \\
{\upshape\ttfamily saueressig@thep.physik.uni-mainz.de} }\\
\end{center}
\vspace{5mm}

\hrule\bigskip

\centerline{\bfseries Abstract} \medskip
\noindent

The gravitational effective average action is studied in a bimetric truncation with a nontrivial background field dependence, and 
its renormalization group flow due to a scalar multiplet coupled to gravity is derived. Neglecting the metric contributions
to the corresponding beta functions, the analysis of its fixed points reveals that, even on the new enlarged theory space which  includes
bimetric action functionals, the theory is asymptotically safe in the large $N$ expansion.
\bigskip
\hrule\bigskip
\newpage
%
\section{Introduction}

The gravitational average action  \cite{mr}  is a universal tool for investigating the scale dependence of the quantum gravitational dynamics.
It can be used in both effective and fundamental field theories of gravity. In particular it has played an important role in the Asymptotic 
Safety program \cite{wein, wein-kall, wein-bern}. In fact, the effective average action seems to evolve along renormalization group (RG)
trajectories  which have exactly the properties postulated by Weinberg \cite{wein}, that is, in the ultraviolet (UV) they run into a nontrivial 
fixed point with a finite dimensional UV-critical surface \cite{mr}-\cite{livrev}. A complete and everywhere regular RG trajectory then defines a 
fundamental and predictive quantum field theory of gravity. At every fixed coarse graining scale the effective average action 
\cite{avact}-\cite{opt} gives rise to an effective field theory valid at that scale. This property has been exploited in some first 
phenomenological investigations of asymptotically safe gravity \cite{bh}-\cite{mof}, 
for instance in inflationary cosmology \cite{entropy}, see also \cite{wein-infl}.

One of the key requirements every future fundamental quantum theory of gravity must meet is that of  ``background independence'' \cite{kiefer}.
Loosely speaking this means that none of the theory's basic rules and assumptions, calculational methods, and none of its predictions, therefore,  
may depend on any special metric that is fixed a priori. All metrics of physical relevance must result from the intrinsic quantum gravitational 
dynamics\footnote{
See \cite{elisa2} and \cite{giulini} for a more detailed discussion of this point.}.
While in loop quantum gravity \cite{A,R,T} and in the discrete approaches \cite{hamber}-\cite{ajl3} the requirement of
``background independence"\footnote{ 
Here and in the following we write ``background independence'' in quotation marks when it is supposed to stand for the above general principle, 
rather than for the independence of a background field.}
is met in the obvious way by completely avoiding the use of any background metric  or a similar non-dynamical structure, this seems very hard to 
do in a continuum field theory. In fact, in the gravitational average action approach \cite{mr} ``background independence'' is implemented by quite 
a different strategy: one introduces an arbitrarily chosen background metric  $\gb_{\m\n}$ at the intermediate steps of the quantization, but verifies at the 
end that no physical prediction depends on which $\gb_{\m\n}$ was chosen. In this way one may take advantage of the entire arsenal of techniques
developed for quantizing fields in a fixed curved background. However, what complicates matters as compared to the usual situation, is that the background spacetime
is never concretely specified; hence there is no way of exploiting the simplifications that would arise for special, highly symmetric backgrounds such as
Minkowski or de Sitter space, say. In a sense, one is always dealing with the ``worst case'' as far as the  complexity of the background structures is concerned.
On the other hand, the crucial advantage of this approach is that it sidesteps all the profound conceptual difficulties and the resulting technical
problems that emerge when one tries to set up a quantum theory without any metric at the fundamental level.  
The difficulties one faces in a program of this type are comparable to those encountered when one tries to quantize a topological field theory 
on a manifold which carries only a smooth but no Riemannian structure.

Thus, from now on we assume that the gravitational degrees of freedom can be encoded in a metric tensor field. We fix a background  metric $\gb_{\m\n}$
and quantize the nonlinear fluctuations of the dynamical metric,  $h_{\m\n}$, in the ``arena'' provided by  $\gb_{\m\n}$, and we repeat this
quantization process for any choice of  $\gb_{\m\n}$. In this manner we arrive at an infinite family of quantum theories for  $h_{\m\n}$, whereby
the family members are labeled by a classical (pseudo-) Riemannian metric $\gb_{\m\n}$. 

The dynamical content of this family is fully described by
an effective action $\G[\hb_{\m\n};\gb_{\m\n}]$ which depends on two arguments\footnote{
For simplicity we ignore the Faddeev-Popov ghosts and possible matter fields here.}:
the expectation value of the fluctuation, $ \hb_{\m\n}\equiv \langle h_{\m\n}\rangle$, and the background metric. If the background-quantum field
split is chosen linear, the expectation value $\hb_{\m\n}$ gives rise to a corresponding expectation value
\be\label{1.1}
g_{\m\n}=\gb_{\m\n}+\hb_{\m\n}
\ee 
of the metric operator $\gamma_{\m\n}$, i.e., $g_{\m\n}=\langle \gamma_{\m\n}\rangle$ where $\gamma_{\m\n}=\gb_{\m\n}+h_{\m\n}$. The action 
$\G[\hb;\gb] $ entails an effective field equation which governs the dynamics of $\hb_{\m\n}(x)\equiv \hb_{\m\n}[\gb](x)$ in dependence
on the background metric:
\be\label{1.2}
\frac{\delta}{\delta \hb_{\m\n}(x)}\G[\hb;\gb]=0.
\ee
For special, so-called ``self-consistent'' backgrounds $\gb_{\m\n}\equiv \gb_{\m\n}^{\textrm{selfcon}}$ it happens that eq.\ \eqref{1.2} is solved by
an identically vanishing fluctuation expectation value $\hb_{\m\n}[\gb_{\m\n}^{\textrm{selfcon}}](x)\=0$. Then the expectation value of the quantum metric 
$g_{\m\n}\equiv g_{\m\n}[\gb]$ equals exactly the background metric, $g_{\m\n}=\gb_{\m\n}$. The defining condition for a self-consistent background,
\be\label{1.3}
\frac{\delta}{\delta \hb_{\m\n}(x)}\G[\hb;\gb^{\textrm{selfcon}}]\Big|_{\hb=0}=0 \, , 
\ee
is referred to as the tadpole equation since it expresses the vanishing of the fluctuation 1-point function. In fact, the corresponding $n$-point
1PI Green's functions for generic $\gb$ are given by 
\be\label{1.4}
\frac{\delta}{\delta \hb_{\m\n}(x_1)}\cdots\frac{\delta}{\delta \hb_{\r\s}(x_n)}\G[\hb;\gb]\Big|_{\hb=0}.
\ee

It is a well known ``magic'' of the background formalism that by using an appropriate gauge fixing condition \cite{back} a set of on-shell
equivalent Green's functions is generated by differentiating the reduced functional $\Gb[\gb]\=\G[\hb;\gb]\Big|_{\hb=0}$ with respect to $\gb_{\m\n}$:
\be\label{1.5}
\frac{\delta}{\delta \gb_{\m\n}(x_1)}\cdots\frac{\delta}{\delta \gb_{\r\s}(x_n)}\Gb[\gb] \, .
\ee
The Green's functions \eqref{1.4} and \eqref{1.5} are equivalent on-shell only if the quantization scheme employed respects the background-quantum field
split symmetry in the physical sector. It expresses the arbitrariness of the decomposition of $g_{\m\n}$ as a background plus a fluctuation.
In the linear case the corresponding symmetry transformation are
\be\label{1.6}
\delta \hb_{\m\n}=\eps_{\m\n},\quad \delta \gb_{\m\n}=-\eps_{\m\n}
\ee
for any $\eps_{\m\n}$. If the split symmetry is appropriately implemented at the quantum level, the reduced functional $\Gb[\gb]$ which depends
on one argument only has exactly the same physical contents as the more complicated $\G[\hb;\gb]$.

Up to now we tacitly assumed that the microscopic  (bare) action governing the dynamics of $\gamma_{\m\n}$ is known a priori, as this would  be the case in an 
ordinary quantum field theory. In the Asymptotic Safety program  \cite{wein}, the situation is different: the pertinent bare action is not an input
but rather a prediction of the theory. More precisely, the idea is to set up a functional coarse graining flow on the space of all
(diffeomorphism invariant) functionals $\G[\hb;\gb]$, to search for nontrivial fixed points of this flow and if there are any, to identify 
the bare action with one of them. This construction yields a quantum field theory with a well behaved UV-limit.

This idea has been implemented in the framework of the gravitational average action \cite{mr, oliverbook}. Here one 
defines a coarse grained counterpart of the ordinary effective action, the effective average action $\G_k[\hb;\gb]$, in terms of a 
functional integral containing an additional mode suppression factor $\exp{(-\Delta_kS)}$ quadratic in $h$, and performs the coarse graining by suppressing the
contributions of the $h_{\m\n}$-modes with a covariant momentum smaller than the variable IR cutoff scale $k$ \cite{avact, ym, avactrev}.
The $k$-dependence of the effective average action  $\G_k[\hb;\gb]$ is governed by a functional RG equation (FRGE) which defines a vector field (a ``flow'') on
theory space. At $k=0$, the average action $\G_k$ equals the ordinary effective action $\G$, and only the gauge fixing term breaks the split symmetry.
Hence, for $k=0$,  $\G_k[\hb;\gb]$  does not contain more gauge invariant information than the reduced functional  $\Gb_k[\gb]\= \G_k[\hb;\gb]\Big|_{\hb=0}$ does.

It is crucial to realize that this (on-shell) equivalence of the two functionals at $k=0$ does {\it not} generalize to $k>0$. At every nonzero scale
$k$ the coarse graining operation unavoidably leads to an additional violation of the split symmetry since the action $\Delta_k S$  is not
invariant. It contains $\hb_{\m\n}$ and $\gb_{\m\n}$ separately, not only in the split invariant combination $\hb_{\m\n}+\gb_{\m\n}$. In a sense,
 $\G_k[\hb;\gb]$  contains more information than $\Gb_k[\gb]$ if $k>0$. An immediate consequence is the well known fact \cite{ym,mr} that it is
impossible to write down a FRGE in terms of the reduced functional $\Gb_k[\gb]$ alone. The actual theory space is more complicated, consisting of
functionals with two metric arguments and, to be precise, also ghost arguments $C^\m$ and $\Cb_\m$, respectively:
$\G_k[\hb_{\m\n},C^\m,\Cb_\m;\gb_{\m\n}]$. Often it is convenient to replace $\hb_{\m\n}$ with $g_{\m\n}\=\gb_{\m\n}+\hb_{\m\n}$ as the  independent 
argument
\be\label{1.7}
\G_k[g,\gb,C,\Cb]\= \G_k[\hb=g-\gb,C,\Cb;\gb] \, .
\ee
In the $\G_k[g,C,\Cb; \gb]$-notation the second argument parametrizes the so-called {\it extra background field dependence}, i.e., that part of the
$\gb_{\m\n}$-dependence which does not combine with a corresponding $\hb_{\m\n}$-dependence to a full metric $\gb_{\m\n}+\hb_{\m\n}\=g_{\m\n}$.
If the split symmetry was intact we had $\tfrac{\delta}{\delta \gb}\G_k[g,\gb,C,\Cb]=0$. In general $\G_k$ has a nontrivial dependence  on $\gb_{\m\n}$
though, and so we can rightfully call $\G_k$ a bimetric action.

The enlarged theory space is the price one has to pay for the ``background independence'' of the average action approach.
Including a matter field $A(x)$ its ``points'' are functionals $\G[A, g,\gb,C,\Cb]$ which are invariant under arbitrary diffeomorphisms acting 
on all arguments simultaneously. If $v^\m$ is a generating vector field and $\cL_v$ the corresponding Lie derivative we have, to first order
in $v^\m$, $\G[A+\cL_v A,g+\cL_v g, \gb+\cL_v \gb,\cdots]=\G[A,g,\gb,\cdots]$. For later use we write down the resulting Ward identity, 
for simplicity at $C=\Cb=0$ and for a scalar matter field:
\be\label{1.8}
2\,g_{\r\n}\,D_\m\,\frac{\d\G[A,g,\gb]}{\d g_{\m\n}(x)}+2\,\gb_{\r\n}\,\Db_\m \, \frac{\d\G[A,g,\gb]}{\d \gb_{\m\n}(x)}-
(\p_\r A)\,\frac{\d\G[A,g,\gb]}{\d A(x)}=0 \, . 
\ee
 Here and in the following the covariant derivatives $D_\m$ and $\Db_\m$ refer to the
Levi-Civit\`a connections of $g_{\m\n}$ and $\gb_{\m\n}$, respectively.

Up to now almost all applications of the gravitational average action employed a truncated theory space with functionals of the form
$\G[g,\gb,C,\Cb]=\Gb_k[\gb]$+{\it classical gauge fixing and ghost terms}, where $\Gb_k[\gb]\=\G_k[\gb,\gb,0,0]$. Hence in practice one
had to deal with the RG evolution of a single metric functional only, $\Gb_k[\gb]$. In ref. \cite{elisa2} a first example of an RG flow with
a nontrivial bimetric truncation was analyzed, albeit only in  conformally reduced gravity rather than the full fledged theory.
In \cite{elisa2} also a number of conceptual issues related to the bimetric character of the average action approach have been explained; we
refer the reader to this discussion for further details. A similar calculation in a different gravity theory has been performed in \cite{floper}.

The purpose of the present paper is to perform the first investigation of a bimetric RG flow involving the full fledged gravitational field. To be precise, we compute the contribution of scalar matter fields to the beta functions of various Newton- and cosmological constant-like running couplings which parametrize $\G_k[A,g,\gb]$ in certain truncations. The quantum effects originating from the gravitational sector are neglected, which allows to avoid the additional technical complications linked to the gauge-fixing of the theory. The resulting RG flow is simple enough to be investigated analytically in a completely explicit way. This enables a detailed study {\it of the conceptual points} underlying the bimetric truncations by identifying and highlighting the general aspects, which will be central to more sophisticated computations in the future \cite{elisafrankII,alcod}. As a spin-off, we re-examine the Asymptotic Safety conjecture within a large-$N$ approximation \cite{perper1,robertoannph}, which gives the first insights on how important the nontrivial bimetric dependence of $\G_k$ actually is.

The remaining sections of this paper are organized as follows. In Section 2 we analyze the RG behavior of the most general bimetric non-derivative term contained in the average action, $\Sg\,\mY_k(g_{\m\n},\gb_{\m\n})$. Here the fluctuation $\hb_{\m\n}=g_{\m\n}-\gb_{\m\n}$ is not required to be small. In Section 3 we employ a complementary truncation of $\G_k$, based on a systematic $\hb_{\m\n}$-expansion which includes all interaction terms built from up to two derivatives and up to first order in $\hb_{\m\n}$. We derive the RG flow on the corresponding five-dimensional theory space and analyze its fixed point structure. The technical details underlying this calculation are relegated to the Appendix. As an application, the resulting $k$-dependent tadpole equation for self-consistent background geometries is discussed in Section 4. Finally, Section 5 contains our conclusions.

\section{The Running Cosmological Constants}
In the following we consider a multiplet of $n_s$ quantized scalar fields $(A_I)\=A,\;I=1,\cdots,n_s,$ coupled to classical gravity. We quantize the scalars
 by means of a (truncated) functional flow equation, being particularly interested in the gravitational interaction terms induced by
 the quantum effects in the matter sector.
In most parts
of the analysis we take the scalars free, except for their interaction with gravity.
The system is described by an effective average action $\G_k[A,g,\gb]$. Even in this comparatively simple setting where gravity itself is not quantized 
this action is ``bimetric'': it depends on the full metric $g_{\m\n}$ since the scalars couple directly to it, and it also depends on $\gb_{\m\n}$ because
it is the background metric which enters the scalar mode suppression term
$\Delta_kS[A;\gb]\=\half\int\textrm{d}^dx\,\Sgb\,A(x)\,\cR_k[\gb]\,A(x)$ where $\cR_k[\gb]$ is the coarse graining operator \cite{avact, ym, mr}.

\subsection{The functional RG equation}
The FRGE for the quantized scalars  $A(x)$ interacting with classical gravity reads
\be\label{3.20}
\partial_t\G_k[A,g,\gb]=\half\,\Tr\left[ \left(\G_k^{(2)}[A,g,\gb]+\cR[\gb]\right)^{-1}\partial_t\cR_k[\gb]\right].
\ee
As gravity is not quantized here, the trace is over the fluctuations of the scalars only, and the Hessian $ \G_k^{(2)}$ involves functional derivatives
with respect to $A$ only. As usual, $t=\ln{k}$ denotes the ``RG time''.

In the following it will be important to carefully distinguish the volume elements $\Sg$ and $\Sgb$, respectively.
The matrix elements of the  Hessian operator contain two factors of the latter,
\be\label{3.21}
\left(\G_k^{(2)}\right)_{xy}\=\langle x\vert \G_k^{(2)}\vert y\rangle=\frac{1}{\sqrt{\gb(x)}\sqrt{\gb(y)}}\,\frac{\delta^2}{\delta A(x)\delta A(y)}\G_k[A,g,\gb].
\ee
Products of operators such as $\G_k^{(2)}$ are defined in terms of their matrix elements as
 $\big(\cA\,\cB\big)_{xy}=\int\textrm{d}^dz\sqrt{\gb(z)}\, \cA_{xz}\,\cB_{zy} $,  and the position representation of the trace in \eqref{3.20}
reads $ \Tr\,\big( \cA\big)=\int\textrm{d}^dx\sqrt{\gb(x)}\; \cA_{xx}$. Furthermore, if some operator has matrix elements $\cA_{xy}$, we define the associated 
(pseudo) differential operator $\cA^{\textrm{diff-op}}$ by $\big(\cA f\big)_x\= \int\textrm{d}^dy\sqrt{\gb(y)}\; \cA_{xy}\, f_y
=\big(\cA^{\textrm{diff-op}} f \big)(x)$  for every ``column vector'' $f_x\= f(x)$. We shall write $\square \equiv g^{\m\n} D_\m D_\n$ and $\bBox \equiv \gb^{\m\n} \Db_\m \Db_\n$ for the Laplace-Beltrami operators belonging to $g_{\m\n}$ and $\gb_{\m\n}$, respectively.

\subsection{The truncation ansatz}

To start with, we  explore the contents of the FRGE \eqref{3.20} with the following ansatz:
\be \label{3.22}
\G_k[A,g,\gb]=\int\textrm{d}^dx\Sg\left( \half g^{\m\n}\partial_{\m}A\, \partial_{\n}A + \half \mb^2_k\, A^2\right) + 
\tfrac{1}{8\pi G}\int\textrm{d}^dx\Sg\; \mY_k \big(g_{\m\n}(x),\gb_{\m\n}(x) \big).
\ee
Here $A$ always stands for  $\big(A_I\big),\;I=1,\cdots,n_s$. The ansatz \eqref{3.22} is taken to be $O(n_s)$ invariant;
appropriate sums over the index $I$ are  understood ($A^2\=A_IA_I$, etc.).
The first term in \eqref{3.22} is the action of a standard scalar coupled to  the {\it full}  metric $g_{\m\n}$, the last represents a generalization of the
running cosmological constant induced by the scalars. It is assumed to contain all possible  non-derivative terms built from $g_{\m\n}$ and $\gb_{\m\n}$.
For the time being we discard induced derivative terms. In particular the running of Newton's constant $G$ is neglected.
The normalization and the explicit factor of $\Sg$ in the second term on the RHS of \eqref{3.22} are chosen such that the constant term of the
function $\mY_k$ equals the cosmological constant:
\be\label{3.23}
\mY_k(g_{\m\n},\gb_{\m\n})=\L_k + \textrm{``more''}.
\ee

In order to project out the beta function of $\mY_k$ it is sufficient to insert $x$-independent metrics  $g_{\m\n}$ and $\gb_{\m\n}$ into the FRGE. In this
case the Hessian resulting from the ansatz \eqref{3.22} is
\be\label{3.24}
\G_k^{(2)}[A,g,\gb]=\frac{\Sg}{\Sgb}\left( -\square + \mb^2_k\right) \, , 
\ee
where we employed the differential operator notation.
\subsection{The cutoff operator}
Now we come to a crucial step which highlights the role of the background metric: the construction of the cutoff operator. Recalling that $\cR_k$ may depend on the
background metric only, we define it by the requirement that upon adding $\cR_k$ to the Hessian {\it and setting $g=\gb$} the operator $ (-\bBox)$ which then
appears in $\G_k^{(2)}[A,\gb,\gb]$ must get replaced by $-\bBox+k^2R^{(0)}\big(-\bBox/k^2\big)$. Here $R^{(0)}$ is an arbitrary shape function \cite{avact, avactrev}
 with the standard 
properties $R^{(0)}(0)=1$ and $\lim_{z\to\io}R^{(0)}(z)=0$. The condition 
\be\label{3.25}
\Big( \G_k^{(2)}[A,g,\gb]+ \cR_k[\gb]\Big)\Big|_{g=\gb}=\G_k^{(2)}[A,\gb,\gb]\Big|_{-\bBox\,\to\;-\bBox+k^2R^{(0)}(-\bBox/k^2)}
\ee
leads to an operator which at first sight appears familiar,
\be\label{3.26}
\cR_k[\gb]\= \cR_k(-\bBox)=k^2\,R^{(0)}\big(-\bBox/k^2  \big)
\ee
which, however, appears in the FRGE combined with the Hessian {\it for $g_{\m\n}$ different from  $\gb_{\m\n}$}:
\be\label{3.27}
\G_k^{(2)}[A,g,\gb]+\cR_k[\gb]=\frac{\Sg}{\Sgb}\left( -\square +\mb_k^2\right) +k^2\,R^{(0)}\big(-\bBox/k^2  \big).
\ee
Most of the unfamiliar features we are going to find in the following are due to the interplay of the standard cutoff operator $\cR_k$
with a Hessian containing the ratio of the volume elements $\Sg/\Sgb$.
\subsection{The RG equation for $\mY_k$}
The trace of the flow equation is easily evaluated in a standard plane wave basis now. With $m\=\mb_k/k$ we have:
\be\label{3.28}
\Sg\;\p_t\,\mY_k(g_{\m\n},\gb_{\m\n})=8\pi G\, n_s\,k^d\int\frac{\textrm{d}^dq}{(2\pi)^d}\,\frac{R^{(0)}(\gb^{\m\n}q_{\m}q_{\n})- 
\big(\gb^{\m\n}q_{\m}q_{\n}\big)\, {R^{(0)}}'(\gb^{\m\n}q_{\m}q_{\n}) }{(\Sg/\Sgb)(g^{\m\n}q_{\m}q_{\n})+R^{(0)}(\gb^{\m\n}q_{\m}q_{\n})+m^2(\Sg/\Sgb) } \, .
\ee
The integral on the RHS of this flow equation 
 represents a scalar density  which depends on two constant matrices, $g^{\m\n}$ and $\gb^{\m\n}$. 
It cannot be evaluated in closed form. In the following we calculate it for two special cases, namely for $g^{\m\n}$ and $\gb^{\m\n}$ both proportional to
the unit matrix, and by expanding in their difference $g_{\m\n}-\gb_{\m\n}=\hb_{\m\n}$.
\subsection{The volume element truncation}
Now we make a further truncation and restrict  $\mY_k$ to depend on the metrics via their volume elements only: $\mY_k(g_{\m\n},\gb_{\m\n})\=\cP_k(\Sg,\Sgb)$.
In this case, the beta function of $\cP_k$ can be found by inserting two conformally flat metrics
\be\label{3.29}
 g_{\m\n}=L^2 \;\delta_{\m\n} \,,\qquad \gb_{\m\n}=\Lb^2\;\delta_{\m\n}
\ee
into the flow equation \eqref{3.28} and keeping track of the constants $L$ and $\Lb$. One then observes that $\p_t \cP_k$ actually depends on the ratio 
$\Sgb/\Sg=(\Lb/L)^d\=\r^d$ only. Thus, setting $\mY_k(g_{\m\n},\gb_{\m\n})\=\cQ_k(\r^d)$ with $\r\=\Lb/L$, we obtain
\be\label{3.30}
\p_t \cQ_k(\r^d)=\, 16\pi\,v_d\, G \, n_s \,k^d \;\r^d\;\int_0^{\io}\textrm{d}y\,y^{d/2-1}\frac{\Rh^{(0)}(y)-y\Rh^{(0)'}(y)}{y+\Rh^{(0)}(y)+ (m/\r)^2} \, .
\ee
Here, $v_d\=\left[ 2^{d+1}\pi^{d/2}\G(d/2)\right]^{-1}$, $y = \Lb^{-2} \delta^{\m\n} q_\m q_\n $, and 
\be\label{3.31}
\Rh^{(0)}(y)\=\r^{d-2}\;R^{(0)}(y) \, . 
\ee

From now on we employ the ``optimized'' shape function \cite{opt} $R^{(0)}(y)=(1-y)\th(1-y)$, whence
\be\label{3.32}
\p_t \cQ_k(\r^d)=\, 16\pi\,v_d\, G \, n_s \, k^d\; \r^d\;\cJ_d(\a,\m).
\ee
Here
\be\label{3.33}
\a\=\r^{2-d}-1\=(\Lb/L)^{2-d}-1\, ,\qquad \m\=m^2\r^{-d} \, ,
\ee
and
\be\label{3.34}
\cJ_d(\a,\m)\=\int_0^{1}\textrm{d}y\,y^{d/2-1}\left[  1+\m+\a y \right]^{-1} \, . 
\ee
The integrals \eqref{3.34} can easily be computed explicitly. Let us specialize for $d=4$ now. Then
\be\label{3.35}
\cJ_4(\a,\m)=\frac{1}{\a}\left[1-\big(\frac{1+\m}{\a}\big)\ln{\left(1+\frac{\a}{1+\m}\right)}\right]
\ee
and the flow equation reads
\be\label{3.36}
\p_t \cQ_k(\r^4)=\frac{G}{2\pi} \, n_s \, \,k^4\;\r^4 \cJ_4(\a,\m) \, .
\ee

At first, let us neglect the scalar mass, setting $m=0$ and $\m=0$ therefore. Then the RG equation \eqref{3.36} can be integrated trivially:
$\cQ_k=\cQ_{k=0}+ \tfrac{G}{8\pi}\,n_s\,k^4\,\r^4 I_4(\a)$, where we abbreviate $I_4(\a)\=\cJ_4(\a,0)$.
As for the constant of integration, $\cQ_{k=0}$, it is important to recall that the $\gb_{\m\n}$-dependence of $\G_k$ is entirely due to the cutoff 
$\cR_k[\gb]$ in the case at hand as we have no gauge fixing terms. As a result, $\G_k$ and in particular $\mY_k$ must become independent of
$\gb$ at $k=0$ since $\cR_k[\gb]$ vanishes there. This implies that $\cQ_{k=0}$ may not have any $\r$-dependence. In fact, it equals the
``ordinary'' cosmological constant $\L_0$ which multiplies the volume element $\Sg$ in the effective action $\G\=\G_0$:
\be\label{3.37}
\cQ_k(\r^4)=\L_0+\frac{1}{8\pi}\;G\;n_s\;k^4\;\r^4 I_4(\a) \,  .
\ee
Using \eqref{3.37} in the truncation ansatz \eqref{3.22} we obtain the following explicit representation for the non-derivative terms in the average action:
\begin{multline}\label{3.38}
\G_k^{\textrm{\tiny non-deriv}}[A,g,\gb]=\frac{1}{8\pi G}\int\textrm{d}^4x\Sg\;\L_0\\
-\frac{n_s}{64\pi^2} \, k^4 \, \int\textrm{d}^4x\;\frac{\Sgb}{1-\sqrt{\Sg/\Sgb}}\left[1+\half\Big\{1-\sqrt{\Sg/\Sgb}\;\Big\}^{-1}\ln{\left(\frac{\Sg}{\Sgb}\right)} \right].
\end{multline}
This is one of the our main results, and several comments are in order here.
\bigskip

\noindent
{\bf (A)} Obviously the induced non-derivative terms are neither proportional to $\Sg$ nor to $\Sgb$ but rather to a complicated function of their ratio,
times an extra factor of $\Sgb$. (The explicit factor of $\r^4$ on the RHS of \eqref{3.37} has converted the $\Sg$ in the ansatz $\int\Sg\;\mY_k$ to 
a $\Sgb$.) 

\noindent
{\bf (B)} The induced term is regular for any ratio $g/\gb\in(0,\io)$. In fact, writing \eqref{3.38} as 
\be\label{3.39}
\G_k^{\textrm{\tiny non-deriv}}[A,g,\gb]=\frac{\L_0}{8\pi G}\;\int\textrm{d}^4x\Sg\;+\;
\frac{n_s}{64\pi^2} \, k^4 \, \int\textrm{d}^4x\;\Sgb\; I_4\left( \a=(g/\gb)^{1/4}-1\right) \, , 
\ee
and using the expansion
\be\label{3.40}
I_4(\a)=\frac{1}{2}-\frac{1}{3}\a+\frac{1}{4}\a^2+ O(\a^3) \, ,
\ee
we see that actually $\G_k^{\textrm{\tiny non-deriv}}[A,g_{\m\n},\gb_{\m\n}]$ has no singularity at $\a=0$, i.e., at $g=\gb$.

\noindent
{\bf (C)} Let us specialize the result for the regime $g\approx\gb$. This amounts to expanding in the fluctuation variable $\hb_{\m\n}\=g_{\m\n}-\gb_{\m\n}$ or,
equivalently, in the variable $\a$ since
\be\label{3.41}
\a\=(g/\gb)^{1/4}-1=\frac{1}{4}\hb_{\m}^{\m}+\frac{1}{32}(\hb_{\m}^{\m})^2-\frac{1}{8}\hb_{\m \n}\hb^{\m \n}+ O(\hb_{\m \n}^3).
\ee
(Here indices are raised and lowered with $\gb_{\m \n}$.) The essential quantity in \eqref{3.39} is $\Sgb\;I_4(\a)$. The first few terms of its $\a$
expansion read
\be\label{3.42}
\Sgb\;I_4(\a)=\frac{1}{2}\Sgb-\frac{1}{3}\left[\sqrt{\Sgb\Sg}-\Sgb \right]+\frac{1}{4}\left[ \Sg -2\sqrt{\Sgb\Sg}+\Sgb \right]+ O(\a^3) \, . 
\ee
In the $g\approx\gb$ regime there is clearly no preference for $g$-monomials over  $\gb$-monomials or vice versa, so the truncation ansatz cannot be simplified
accordingly: any meaningful truncation is genuinely ``bimetric''! Note also that the monomial $\sqrt{\Sg\Sgb}$ which was individually included in
 the truncation studied in \cite{elisa2}\footnote{
Note, however, that $\sqrt{\Sg\Sgb}$ plays a distinguished role in the conformally reduced gravity setting of \cite{elisa2}. It amounts  to a mass term 
of the scalar field appearing there. }
 never is generated in isolation. According to \eqref{3.42} it is always accompanied by other,
a priori equally important terms involving $\Sg$ and $\Sgb$.

\noindent
{\bf (D)} It is instructive to rewrite the expansion about $g=\gb$ in terms of $\hb_{\m\n}$. Up to linear order the relevant terms in 
$\G_k[A,g_{\m\n},\gb_{\m\n}]\=\G_k[A,\hb_{\m\n};\gb_{\m\n}]$ are
\be\label{3.43}
\G_k^{\textrm{\tiny non-deriv}}[A,\hb;\gb]=\frac{\L_k^{(0)}}{8\pi G} \int \textrm{d}^4x\Sgb\;+\;\frac{\L_k^{(1)}}{8\pi G}\cdot\frac{1}{2} \int \textrm{d}^4x\Sgb\;\gb^{\m\n}\hb_{\m\n}
+ O(\hb_{\m\n}^2).
\ee
Here $\L_k^{(0)}$ is a {\it background cosmological constant} multiplying $\int\Sgb$ rather than $\int\Sg$, and $\L_k^{(1)}$ is an analogous, but
numerically different running coupling in the (only) term linear in $\hb_{\m\n}$. Explicitly,
\be
\L_k^{(0)} = \L_0+\frac{n_s}{16\pi}\;G\;k^4 \, , \qquad
\L_k^{(1)} = \L_0-\frac{n_s}{48\pi}\;G\;k^4 \, . \label{3.44}
\ee
Note the different signs on the RHS of the equations: $\L_k^{(0)}$ increases, but $\L_k^{(1)}$ decreases for growing $k$. Note also that when $k\neq 0$
the RHS of \eqref{3.43} cannot be written as a functional of the single metric $\gb_{\m\n}+ \hb_{\m\n}\=g_{\m\n}$ alone as this would require it to be 
proportional to $\Sg=\Sgb+\half\Sgb\;\gb^{\m\n}\hb_{\m\n}+ O(\hb_{\m\n}^2)$. But this is not the case just because  $\L_k^{(0)}\neq \L_k^{(1)}$ .

\noindent
{\bf (E)} Computations within the setting of single metric truncations retain only the terms of zeroth order in $ \hb_{\m\n}$. 
They equate the two metrics $g_{\m\n}$ and $\gb_{\m\n}$, and traditionally denote the one metric which is left over then by $g_{\m\n}$. 
In this setting, eq.\ \eqref{3.43} would boil down to
\be\label{3.46} 
\G_k^{\textrm{\tiny non-deriv,single metric}}[A,g,g]=\frac{\L^{(0)}_k}{8\pi G}\int \textrm{d}^4x\Sg \, .
\ee
Thus, in a single metric truncation, it is the parameter  $\L^{(0)}_k$ which would be interpreted as ``the'' cosmological constant, the one, and  only one, responsible for 
the curvature of spacetime. From the more general perspective of the bimetric truncation we understand that this is actually missleading. We shall discuss this
point in detail in Section \ref{sect.4}.

\noindent
{\bf (F)} We saw that, in the regime $\Sg\approx\Sgb$, the true cosmological constant monomial $\Sg$ plays no distinguished role. Next let us see whether
this could be the case when $\Sg\ll \Sgb$. Let us try an expansion in $\s^2\=\sqrt{\Sg/\Sgb}\ll1$. In a $\s^2$-expansion of \eqref{3.39} with $\a\=\s^2-1$
there would indeed appear a $\Sg$ monomial in isolation, the first few terms of the power series being $\Sgb(\s^2)^0=\Sgb,\quad \Sgb(\s^2)^1=\sqrt{\Sg\Sgb},
\quad\Sgb(\s^2)^2=\Sg,\quad\Sgb(\s^2)^3=g^{3/4}\gb^{-1/4},\cdots$. However, taking the explicit form of $I_4$ into account one finds that an expansion of 
this type actually does not exist, since $I_4(\a=\s^2-1)\approx -(1+\ln{\s^2})$ when $\s\ll1$. Hence $I_4$ is not analytic at $\s^2=0$. 
So the conclusion is that in this regime, too, the $\Sg$-monomial plays no  distinguished role in the truncation. (The same is true for the $\sqrt{\Sg\Sgb}$
term considered in \cite{elisa2}.)

\noindent
{\bf (G)} Up to now, the scalars $A$ were assumed massless. If we  allow for a non-zero dimensionless mass $m\equiv\mb_k/k$ so that now
$\m=(\mb^2_k/k^2)(\Sg/\Sgb)^d$, the RG equation \eqref{3.36} no longer can be integrated trivially. The general properties of the flow are clear though:
if $k\gg \mb_k$, the matter field is still approximately massless and the above discussion applies. If, on the other hand, $k\ll\mb_k$ the scalar decouples
and its contribution to the running of $\cQ_k$ becomes tiny. If we neglect the running of the dimensionful mass  $\mb_k\=M$ the parameter $\m$
diverges $ \propto 1/k^2$ below the  threshold at $k=M$. Expanding $\cJ(\a,\m)=\tfrac{1}{2\m}+O(\tfrac{1}{\m^2})$ we obtain  the leading term below 
the threshold:
\be\label{3.47}
\G_k^{\textrm{\tiny non-deriv}}=\frac{1}{8\pi G}\int\textrm{d}^4x\left\{ \Sg\L_0+\Sgb\left(\frac{\Sgb}{\Sg}\right)\frac{G}{24\pi}\,n_s\,k^4 \, \left[ \frac{k^2}{M^2}+
O\left( \frac{k^4}{M^4}\right)\right]\right\}.
\ee
In the massive regime an extra factor $k^2/M^2\ll1$ suppresses the running of the non-derivative term. Here, again, it is not of  the standard $\Sg$-form,
but rather proportional to $\gb/\Sg$.

\noindent
{\bf (H)} So far, the discussion was based upon an evaluation of the integral formula \eqref{3.28} for metrics $g_{\m\n}$ and $\gb_{\m\n}$
which are conformal to $\delta_{\m\n}$, see \eqref{3.29}. If one wants to know the tensorial structure of the terms in $\mY_k$ one must go beyond this
special case. A systematic strategy for doing this is an expansion in $\hb_{\m\n}=g_{\m\n}-\gb_{\m\n}$, whereby $g_{\m\n}$ and $\gb_{\m\n}$
are still constant matrices, but  not necessarily proportional to $\delta_{\m\n}$. Carrying out this expansion up to linear order in $\hb_{\m\n}$ we find a
result of the form \eqref{3.43}, generalized for arbitrary $d$, with
\be\label{3.48}
\begin{split}
\L_k^{(0)} = & \, \L_0+\frac{n_s}{(4\pi)^{d/2-1}}\;\frac{2}{d} \;\Phi^1_{d/2}(0)\;G\;k^d \, , \\
\L_k^{(1)} = & \, \L_0-\frac{n_s}{(4\pi)^{d/2-1}}\;\frac{(d-2)}{d}\; \Phi^2_{d/2+1}(0)\;G\;k^d \, .
\end{split}
\ee
Here the $\Phi$'s are the usual $R^{(0)}$-dependent threshold function of ref. \cite{mr}, see also eq.\ \eqref{A.18} in the Appendix.
 If one specializes for $d=4$ and the ``optimized'' 
$R^{(0)}$, the result \eqref{3.48} accidentally coincides with the one obtained from the conformally flat ansatz \eqref{3.44}. As \eqref{3.48} 
 contains different $\Phi$-functions, they depend on  $R^{(0)}$ in a different
way. This implies that any special relationship between the two cosmological constants  $\L_k^{(0)}$ and $\L_k^{(1)}$ which one might invoke,
for instance in order to restore split symmetry,
cannot have a universal (scheme independent) meaning at $k\neq 0$.

\section{ A systematic derivative- and $\hb_{\m\n}$-expansion}
\label{sect.3}

In this section we explore the matter induced gravitational coupling constants in a different parameterization of the average action. It comprises the 
first terms of a systematic expansion of $\G_k[A,g,\gb]\=\G_k[A,\hb;\gb]$ in powers of $\hb_{\m\n}$ and the number of derivatives.

\subsection{The truncation ansatz} 
In the following, we will consider an average action of the form
\be\label{4.1}
\begin{split}
\G_k[ A,\hb;\gb]= & \, 
-\frac{1}{16\pi G_k^{(0)}} \int\textrm{d}^dx\Sgb\left(\Rb-2\L_k^{(0)} \right)\\
& \, + \frac{1}{16\pi G_k^{(1)}} \int\textrm{d}^dx\Sgb\left(\Gbb^{\m\n}-\half\, E_k\;\gb^{\m\n}\;\Rb+ \L_k^{(1)}\;\gb^{\m\n} \right)\hb_{\m\n}\\
& \, + \int\textrm{d}^dx\Sg\left( \half g^{\m\n}\;\partial_{\m}A\, \partial_{\n}A + \cU_k(A) \right)\Big|_{g_{\m\n}=\gb_{\m\n}+\hb_{\m\n}} \, .
\end{split}
\ee
Here $\Rb$ is the curvature scalar built from the background metric, $\Gbb^{\m\n}\=\gb^{\m\r}\gb^{\n\s}\Gbb_{\r\s}$, 
and $\Gbb_{\m\n}\=\Rb_{\m\n}-\half\gb_{\m\n}\,\Rb$ denotes the background Einstein tensor.
The functional \eqref{4.1} is obviously  invariant under diffeomorphisms acting simultaneously on $\hb_{\m\n},\;A$, and $\gb_{\m\n}$, respectively.
The gravitational part of this ansatz is complete in the sense that it contains all possible terms with no or one factor of  $\hb_{\m\n}$ and at most
two derivatives. The latter can always be arranged to act on $\gb_{\m\n}$, and  diffeomorphism invariance then implies that they occur as 
contractions of the background Riemann tensor. 
 The matter part of \eqref{4.1} has the same structure as in the previous section; now $g_{\m\n}$ is to be read as an abbreviation for
$\gb_{\m\n}+\hb_{\m\n}$, though.

There exist two $\hb_{\m\n}$-independent field monomials with zero and two derivatives, respectively, namely $\int\Sgb$ and $\int\Sgb\,\Rb$. In eq. \eqref{4.1}
the corresponding prefactors are proportional to $\L^{(0)}_k/G_k^{(0)}$ and $1/G_k^{(0)}$, respectively. In the sector with one power of 
 $\hb_{\m\n}$ there are three possible tensors structures for the corresponding one-point function, namely $\Gbb^{\m\n}$ and $\gb^{\m\n}\Rb$ with two, and 
$\gb^{\m\n}$ with no derivatives. Their prefactors define new running couplings $1/G_k^{(1)}$,  $E_k/G_k^{(1)}$, and $\L^{(1)}_k/G_k^{(1)}$, respectively.
(The superscripts $(0),(1),\dots$ indicate the $\hb_{\m\n}$-order in which the coupling in question occurs.)

The functional \eqref{4.1} is defined for completely independent fields $\hb_{\m\n}$ and $\gb_{\m\n}$; hence $\G_k$ has an ``extra'' $\gb_{\m\n}$-dependence in 
general which does not combine with $\hb_{\m\n}$ to a full metric $g_{\m\n}=\gb_{\m\n}+ \hb_{\m\n}$. Nevertheless it is instructive to consider 
\eqref{4.1} for the special case of no extra $\gb_{\m\n}$-dependence. The background-quantum field split symmetry is  intact then and $\G_k$ depends on
$\gb_{\m\n}$ and $\hb_{\m\n}$ via their sum only. The resulting gravitational part is obtained by expanding the Einstein-Hilbert action,
\be\label{4.2}
\Gamma_k^{\rm EH}[\gb+\hb] \= -\frac{1}{16\pi G_k^{(0)}} \int\textrm{d}^dx\Sg\left(R(g)-2\L_k^{(0)} \right)\Big|_{g_{\m\n}=\gb_{\m\n}+\hb_{\m\n}} \, ,
\ee
up to terms of first order in $\hb_{\m\n}$. The expansion is of the form \eqref{4.1} with special coefficients, 
however:
\begin{subequations}\label{4.3}
\begin{align}
G_k^{(1)}= &\;\; G_k^{(0)},\label{4.3a}\\
\L_k^{(1)}=&\;\;\L_k^{(0)},\label{4.3b}\\
E_k=& \;\;0\label{4.3c}.
\end{align}
\end{subequations}

It needs to be stressed that the relations \eqref{4.3} are {\it not} satisfied in general. Since the cutoff term $\Delta_k S$ breaks the split symmetry,
$\G_k$ does have an extra background dependence, and so the terms  with one power of $\hb_{\m\n}$ are not the linearization of any functional
depending on the sum $\gb_{\m\n}+ \hb_{\m\n}$ only. As a consequence, we encounter two running couplings, $ G_k^{(0)}$ and $ G_k^{(1)}$, both of which
are related to the classical Newton constant, but have a different conceptual status and numerical value. The same remark applies to the running cosmological constants 
$\L_k^{(0)}$ and $\L_k^{(1)}$.

 Furthermore,  we emphasize that when the split symmetry is broken there is no symmetry principle that would force
the second derivative terms with one  $\hb_{\m\n}$ factor to be proportional to the Einstein tensor. There are actually two independent tensor structures,
$\Rb^{\m\n}\,\hb_{\m\n}$ and $\Rb\, \gb^{\m\n}\,\hb_{\m\n}$, respectively; they can equivalently be chosen as $\Gbb^{\m\n}\,\hb_{\m\n}$ together with 
 $\Rb\, \gb^{\m\n}\,\hb_{\m\n}$.
Hence $E_k\neq0$ in general. (The implications of a non-vanishing $E_k$ for the effective field equations will be discussed below.)

If, as in the system discussed in the present paper, the cutoff term $\Delta_k S$ is the only source of split symmetry violation the average action looses its
extra $\gb$-dependence in the ``physical  limit''$k\to0$. We expect the relations \eqref{4.3} to be satisfied then. In non-gauge theories the ordinary 
effective action $\lim_{k\to0} \G_k[A,g,\gb]\=\G[A,g]$ has no ``extra'' $\gb$-dependence.

\subsection{The five-dimensional RG flow}
In order to obtain the beta functions for the running couplings in the truncation ansatz we must insert \eqref{4.1} into the flow equation \eqref{3.20},
compute the trace on its RHS, and project it on the various monomials. For bimetric truncations computations of this kind are considerably more 
involved than in the single metric case. In the present situation where one must retain only one power of $\hb_{\m\n}$ the calculation still can be done in 
a comparatively elegant way. Some details, in particular the new strategies to cope with the $\hb_{\m\n}$-dependence are given in the Appendix.
 Here we only present and analyze the results.

As we are not interested in the renormalization of the matter sector we only include a mass term in the potential: $\cU_k(A)=\half\mb_k^2 A^2$. The 
dimensionful mass $\mb_k$ has the same value for all components of $A$, and we neglect its scale dependence, $\mb_k\= \mb$. The RG equations
are most conveniently written down in terms of the dimensionless quantities $m_k\=\mb/k,\, E_k,$ and 
\be\label{4.4}
g_k^{(0)}\=k^{d-2}\;G_k^{(0)}\, ,\qquad  \l_k^{(0)}\=k^{-2}\L_k^{(0)} \, , \qquad
g_k^{(1)}\=k^{d-2}\;G_k^{(1)}\, ,\qquad  \l_k^{(1)}\=k^{-2}\L_k^{(1)} \, .
\ee
Furthermore, it is convenient to introduce the following anomalous dimensions for the two running Newton constants:
\be\label{4.5}
\eta_N^{(0)}\=\frac{\p_t G_k^{(0)}}{G_k^{(0)}}\quad,\quad \eta_N^{(1)}\=\frac{\p_t G_k^{(1)}}{G_k^{(1)}}.
\ee

For the couplings of order zero in $\hb_{\m\n}$, the RG equations assume the following form:
\begin{subequations}\label{4.6}
\begin{align}
 \p_t \,g_k^{(0)}= & \left[ d-2+\eta_N^{(0)}\right]\,g_k^{(0)}\label{4.6a} \, , \\
\p_t\,\l_k^{(0)}= & \left[ \eta_N^{(0)}-2\right]\,\l_k^{(0)}\,+\,2\,n_s\,(4\pi)^{1-d/2}\,g_k^{(0)}\; \Phi^1_{d/2}(m_k^2)\label{4.6b} \, .
\end{align}
 \end{subequations}
Similarly, one finds at order one:
\begin{subequations}\label{4.7}
\begin{align}
 \p_t \,g_k^{(1)}= & \left[ d-2+\eta_N^{(1)}\right]\,g_k^{(1)}\label{4.7a} \, , \\
\p_t\,\l_k^{(1)}= & \left[ \eta_N^{(1)}-2\right]\,\l_k^{(1)}\,-
\,n_s\,(4\pi)^{1-d/2}\,g_k^{(1)}\,\left[ (d-2)\,\Phi^2_{d/2+1}(m_k^2)+2m_k^2\; \Phi^2_{d/2}(m_k^2)\right]
\label{4.7b} \, , \\
\p_t\,E_k= &\left[\half(d-2)+E_k\right] \eta_N^{(1)}+\tfrac{2}{3}\,n_s\,(4\pi)^{1-d/2}\,m_k^2\;g_k^{(1)}\; \Phi^2_{d/2-1}(m_k^2)\label{4.7c}.
\end{align}
 \end{subequations}
The anomalous dimensions are explicitly given by
\begin{subequations}\label{4.8}
\begin{align}
\eta_N^{(0)}= & \frac{2}{3} \,n_s\,(4\pi)^{1-d/2}\,g_k^{(0)}\; \Phi^1_{d/2-1}(m_k^2)\label{4.8a},\\
\eta_N^{(1)}= & \frac{2}{3} \,n_s\,(4\pi)^{1-d/2}\,g_k^{(1)}\; \Phi^2_{d/2}(m_k^2)\label{4.8b}.
\end{align}
 \end{subequations}

Various remarks are in order here:
\bigskip

\noindent
{\bf (A)} The above five ordinary differential equations for $g_k^{(0,1)}$, $\l_k^{(0,1)}$, and $E_k$ form two sets which are completely decoupled: 
the order zero quantities  $g_k^{(0)}$ and $\l_k^{(0)}$ enter only the coupled system \eqref{4.6}, while the order-one couplings
$g_k^{(1)}$, $\l_k^{(1)}$, and $E_k$ occur only in the three-dimensional system \eqref{4.7}.\footnote{The complete decoupling of the two sets of gravitational flow equations is accidental, however, and owed to the simplicity of the model. In a more
elaborate framework, which also includes the quantum effects in the gravitational sector, the coupling constants appearing at higher orders in the $\hb$-expansion will, in general, ``feed back'' into the beta functions capturing the running of the coupling constants at lower orders \cite{elisa2,elisafrankII}.}

\noindent
{\bf(B)} The two-dimensional subsystem for the order zero quantities, eqs.\ \eqref{4.6a} and \eqref{4.6b}, is exactly what one obtains in a
conventional single metric truncation, with   $g_k^{(0)}=g_k^{(1)}$ and $\l_k^{(0)}=\l_k^{(1)}$ interpreted  as ``the''
Newton and cosmological constant, respectively, \cite{robertoannph}.

\noindent
{\bf(C)} The above equations make it manifest that the RG flow does indeed generate split symmetry violating terms: the beta function of $E_k$
is nonzero, and the beta functions for the zeroth and first order Newton and cosmological constants are different.

\noindent
{\bf(D)} As we want split symmetry to be restored at $k=0$, the constants of integration contained in solutions to the above RG equations must
be chosen in accord with \eqref{4.3}. This does not fix them completely, of course, but reduces the set of undetermined coupling constants to the one
arising in the single metric computation.

\subsection{Nontrivial fixed points in the large $N$ approximation}
We close this section by studing the fixed point structure of the RG flow defined by the above equations.
In the large $N$ limit $N\=n_s\to\io$ the renormalization effects due to the matter fields overwhelm those stemming from the gravitational field
which have been neglected here. Hence 
 it makes sense to ask whether the above RG equations admit ``asymptotically safe'' solutions, i.e. RG trajectories which run into a non-Gaussian
fixed point (NGFP) when $k\to\io$.

For the sake of simplicity we restrict the discussion to $d=4$ and $m=0$. 
The system for the order-zero quantities \eqref{4.6} then becomes
\begin{subequations}\label{4.20}
\begin{align}
 \p_t \,g_k^{(0)}= & \left[ 2+\eta_N^{(0)}\right]\,g_k^{(0)}\quad ,\quad \eta_N^{(0)}=  \frac{n_s}{6\pi} \,g_k^{(0)}\; \Phi^1_{1}(0),\label{4.20a}\\
\p_t\,\l_k^{(0)}= & \left[ \eta_N^{(0)}-2\right]\,\l_k^{(0)}\,+ \frac{n_s}{2\pi}\,g_k^{(0)}\; \Phi^1_{2}(0).\label{4.20b}
\end{align}
 \end{subequations}
The three coupled equations for the order-one couplings \eqref{4.7} simplify to:
\begin{subequations}\label{4.21}
\begin{align}
 \p_t \,g_k^{(1)}= & \left[ 2+\eta_N^{(1)}\right]\,g_k^{(1)}\quad,\quad \eta_N^{(1)}=  \frac{n_s}{6\pi} \,g_k^{(1)}\; \Phi^2_{2}(0),
\label{4.21a}\\
\p_t\,\l_k^{(1)}= & \left[ \eta_N^{(1)}-2\right]\,\l_k^{(1)}\,-\,\frac{n_s}{2\pi}\,g_k^{(1)}\; \Phi^2_{3}(0),
\label{4.21b}\\
\p_t\,E_k= & \left[1 +E_k\right] \eta_N^{(1)}.\label{4.21c}
\end{align}
 \end{subequations}
Owed to their decoupling, the two sets of equations can then be analyzed independently.

It is easy to see that both the order zero and the order one subsystem has a Gaussian fixed point (GFP) as well as a non-Gaussian fixed point (NGFP). From \eqref{4.20}
we read off that the beta functions of $g_k^{(0)}$  and $\l_k^{(0)}$ vanish at the following  two points:
\begin{subequations}\label{4.22}
 \begin{align}
  \textrm{GFP}^{(0)}:& {}&  g_*^{(0)}& =0, &\l_*^{(0)}&=0\label{4.22a} \, , \\
  \textrm{NGFP}^{(0)}:&{}&   g_*^{(0)}& =-\frac{12 \pi}{n_s\, \Phi^1_1(0)},& \l_*^{(0)}&=-\frac{3}{2}\;\frac{\Phi^1_2(0)}{\Phi^1_1(0)}\label{4.22b} \, . 
 \end{align}
\end{subequations}
The zeros of the order one beta functions are:
\begin{subequations}\label{4.23}
 \begin{align}
  \textrm{GFP}^{(1)}: & {} & g_*^{(1)} &=0,  &\l_*^{(1)}&=0,  & E_* & \textrm{  arbitrary} \label{4.23a} \, , \\
  \textrm{NGFP}^{(1)}: &{} & g_*^{(1)} &=-\frac{12 \pi}{n_s\, \Phi^2_2(0)}, &\l_*^{(1)}&=\frac{3}{2}\;\frac{\Phi^2_3(0)}{\Phi^2_2(0)}, & E_*&=-1 \, .
\label{4.23b}
 \end{align}
\end{subequations}

The following points should be noted here.

\noindent
{\bf (A)} Any fixed point of the order zero system may be combined with any fixed point of the order one equations. Hence we find 
four fixed points of the total system; symbolically:
\be\label{4.24}
\textrm{GFP}^{(0)}\otimes\textrm{GFP}^{(1)},\quad \textrm{GFP}^{(0)}\otimes \textrm{NGFP}^{(1)},\quad  \textrm{NGFP}^{(0)}\otimes \textrm{GFP}^{(1)},\quad 
 \textrm{NGFP}^{(0)}\otimes \textrm{NGFP}^{(1)} \, .
\ee

\noindent
{\bf (B)} The constants $\Phi^p_n(0)$ are cutoff scheme, i.e., $R^{(0)}$-dependent, but their signs are universal. At both non-Gaussian fixed points the respective
Newton constants assume {\it negative} values. (In the case of $ g_*^{(0)}$ this was known already \cite{robertoannph}.) Inserting the ``optimized''
shape function as an example one finds:
\begin{subequations}\label{4.25}
 \begin{align}
  \textrm{NGFP}^{(0)}:&{} & g_*^{(0)}&=-\frac{12\pi}{n_s},& \l_*^{(0)}&=-\frac{3}{4} \, ,  &{}\label{4.25a} \\
  \textrm{NGFP}^{(1)}:&{} & g_*^{(1)}&=-\frac{24\pi}{n_s},& \l_*^{(1)}&=\frac{1}{2},& E_*=-1\label{4.25b} \, . 
\end{align}
\end{subequations}
Note also the different signs of $\l_*^{(0)}$ and  $\l_*^{(1)}$.

\noindent
{\bf (C)} While we knew already from the single metric truncations that there exists a NGFP in the order zero sector, the result
concerning a nontrivial fixed point at order one is new. Clearly this is encouraging news for the Asymptotic Safety program. It requires 
a fixed point on the ``big'' theory space spanned by functionals of the type $\G[A,\hb;\gb]$, and they comprise terms of all orders in $\hb_{\m\n}$, of course.

\noindent
{\bf (D)} If there was no split symmetry breaking the fixed points in the zero and first order  sectors were related, with \eqref{4.3} implying
\begin{subequations}\label{4.26}
 \begin{align}
   g_*^{(1)} & = g_*^{(0)}\label{4.26a} \, , \\
\l_*^{(1)} & = \l_*^{(0)}\label{4.26b} \, , \\
E_* & = 0 \label{4.26c} \, . 
 \end{align}
\end{subequations}
Clearly the explicit results \eqref{4.25} are far from satisfying these relations, and we must conclude that the violation of the split symmetry
due to the mode cutoff has a significant impact on the fixed point structure. It needs to be emphasized that it is quite nontrivial
 that the known single metric fixed point did not get destroyed within the more general bimetric truncation. We rather find that it splits
into two different ones, one at the zeroth and another one at the  first  level of the $\hb_{\m\n}$-expansion.

\noindent
{\bf (E)} The  running coupling $E_k$ enters the  action \eqref{4.1} via the combination 
\be\label{4.27}
\Gbb^{\m\n}-\half\,E_k\, \gb^{\m\n}\;\Rb\,=\, \Rb^{\m\n}-\half(1+E_k)\,\gb^{\m\n}\;\Rb \, .
\ee
The $\textrm{NGFP}^{(1)}$ fixed point value  $E_*=-1$ is quite special therefore. It  is precisely 
such that  the Einstein tensor is converted  to a pure Ricci tensor. Note that also that at the $\textrm{GFP}^{(1)}$ the value of $E_*$
is arbitrary: if $g_k^{(1)}=0$, the RHS of \eqref{4.21c} vanishes whatever is the value of $E_k$. So strictly speaking we encounter a
whole line of fixed points. Imposing split symmetry leads to $E_*=0$ though.

\noindent
{\bf (F)} Imposing the restauration of split symmetry at $k=0$, it is possible to solve the RG equation \eqref{4.21c}  explicitly for $E_k$ in terms of $ G_k^{(1)}$, for all values of $k$
\be\label{4.28}
E_k= G_k^{(1)} /G_0-1 \, .
\ee
Here $ G_0$ is the common value of $ G_k^{(0)}$ and $G_k^{(1)}$ at $k=0$. In the semiclassical regime we have approximately 
\mbox{$G_k^{(1)}=G_0\left[1+\tfrac{n_s}{12\pi} \Phi^2_2(0)\,G_0\,k^2 \right]$} and $E_k=\tfrac{n_s}{12\pi}\Phi^2_2(0)\,G_0\,k^2 $, hence
$E_k$ is a tiny number of order $k^2/m_{\textrm{Pl}}^2$ there.

\noindent
{\bf (G)} The negative sign of $ g_*^{(0)}$ and $ g_*^{(1)}$ is of no relevance to the formal considerations of the present paper; actually $ g_*^{(0)}$
is known to be positive for fermionic matter fields \cite{robertoannph}.

\section{Tadpole equation and self-consistent backgrounds}
\label{sect.4}
%
One of the perhaps somewhat unusual features of the effective average action at $k\neq0$ is that its field-source relations (``effective field equations'')
which govern the expectation value fields involve $\tG_k\=\G_k+\Delta_k S$ rather than  $\G_k$. In an arbitrary theory with
dynamical (i.e., non-background) fields $\Phi_i$ these relations read, for vanishing sources \cite{avact},
\be\label{4.30}
\frac{\d \tG_k[\Phi]}{\d\Phi_i}\equiv \frac{\d \G_k[\Phi]}{\d\Phi_i}+\frac{\d \Delta_k S[\Phi]}{\d\Phi_i}=0 \, .  
\ee
Since, symbolically, $\Delta_k S\propto \int \Phi\,\cR_k\,\Phi$, the last term in eq.\ \eqref{4.30} equals essentially $\cR_k\Phi_i$. It
vanishes at $k=0$ since $\cR_{0}=0$ by construction.

For the gravity-scalar system of the present paper the set ($\Phi_i$) comprises $A$ and $\hb_{\m\n}$. The effective field equation for the scalar reads
\be\label{4.31}
\frac{\d \G_k[A,\hb;\gb]}{\d A(x)}+\Sgb\,\cR_k[\gb]\,A(x)=0 \, . 
\ee
A similar equation, involving a term $\cR_k\,\hb_{\m\n}$, holds for the metric fluctuation. This latter equation simplifies when we search 
for self-consistent backgrounds, since then we impose $\hb_{\m\n}=0$ so that the  $\cR_k\,\hb_{\m\n}$ term is absent. Hence
the generalization of the tadpole equation \eqref{1.3} for $k\neq 0$ reads simply
\be\label{4.32}
\frac{\d}{\d \hb_{\m\n}(x)}\G_k[A,\hb;\gb^{\textrm{selfcon}}]\Big|_{\hb=0}=0 \, . 
\ee
We call $\gb^{\textrm{selfcon}}$ a self-consistent background metric for the scale $k$ if there exists a scalar field configuration $A(x)$ such that the 
coupled system of equations \eqref{4.31} and \eqref{4.32} is satisfied.

For the truncation ansatz \eqref{4.1} the scale dependent equation \eqref{4.32} reads explicitly, omitting the superscript from $\gb^{\textrm{selfcon}}$,
\be\label{4.33}
\Gbb^{\m\n}-\half\,E_k\,\gb^{\m\n}\,\Rb+\L_k^{(1)}\,\gb^{\m\n}=8\pi\,G_k^{(1)}\;T^{\m\n}[A,\gb] \, . 
\ee
Here  $T^{\m\n}$ denotes the Euclidean energy-momentum tensor
\be\label{4.34}
T^{\m\n}[A,\gb]\equiv -2 \, \frac{1}{\Sgb}\,\frac{\d}{\d g_{\m\n}}\G_k^{\textrm{M}}[A,g,\gb]\Bigg|_{g=\gb} \,  .
\ee
Within the truncation studied above the matter part of the average action, $\G_k^{\textrm{M}}[A,g,\gb]$, is given by the last line in eq.\ \eqref{4.1}.
In this section we are slightly more general and allow for an arbitrary matter action which may have an ``extra'' $\gb$-dependence.
Above, such a dependence was excluded by the form of the ansatz, but in more general truncations an ``extra'' $\gb$-dependence 
can be induced by the RG running. At $k\neq0$ gravity interacts with matter not only via the full metric  $g=\gb+\hb$,
but also by split symmetry violating ``extra'' $\gb$-couplings.

The condition for a self-consistent background \eqref{4.33} differs from Einstein's equation by the $E_k$-term on its LHS. This term
is not forbidden by any general principle since \eqref{4.33} is {\it not} the variation of a diffeomorphism invariant functional $F[A,g]$,
evaluated at $g=\gb$. If it was, the pure gravity part in the 2-derivative sector of $F[A,g]$ could only be $\int\Sg\,R(g)$, which produces
a pure Einstein tensor with no $E_k$-correction.
In fact, the $E_k$-term owes its existence to the violated split symmetry or, stated differently, to the bimetric character of
$ \G_k[A,g,\gb]\=\G_k[A,\hb;\gb]$. What goes beyond the familiar variational principle that would lead to the standard Einstein equation is
precisely the ``extra'' $\gb$-dependence of $\G_k[A,g,\gb]$.

Interestingly, the tadpole equation \eqref{4.33} can be written as an ordinary Einstein equation, with no $E_k$-term on its LHS, but
with a modified energy momentum tensor on its RHS. To see this we take the trace\footnote{
For simplicity we set $d=4$ here. }
of \eqref{4.33}
\be\label{4.35}
\Rb=\frac{1}{1+2E_k}\Bigg(4\L_k^{(1)} -8\pi\,G_k^{(1)} \,\gb_{\m\n}\,T^{\m\n}\Bigg) \, . 
\ee 
If we substitute this formula for the curvature scalar into the $E_k$-term of \eqref{4.33} we arrive at a conventionally-looking Einstein equation
\be\label{4.36}
\Gbb^{\m\n}= -\widetilde{\L}_k^{(1)}\,\gb^{\m\n}+8\pi\,G_k^{(1)}\, \widetilde{T}^{\m\n}[A,\gb] \, . 
\ee
Yet, this equation contains a rescaled cosmological constant,
\be\label{4.37}
\widetilde{\L}_k^{(1)}\=\frac{\L_k^{(1)}}{1+2E_k} \, ,
\ee
and a modified energy-momentum tensor
\be\label{4.38}
\widetilde{T}^{\m\n}[A,\gb]\= T^{\m\n}[A,\gb]-\frac{E_k}{2(1+2E_k)}\,\gb^{\m\n}\gb_{\a\b}\, T^{\a\b}[A,\gb] \, .
\ee 

The ordinary Einstein equation comes  with a nontrivial integrability condition which is of central importance, both
from the mathematical and the physical point of view. As a consequence of Bianchi's identity, the equation is consistent only
when the energy-momentum tensor is covariantly conserved. For the tadpole equation \eqref{4.33} the situation is more complicated.
Applying $\Db_\m$ to it we obtain
\be\label{4.39}
\Db_\m  T^{\m\n}[A,\gb]=-\frac{1}{2}\Big(8\pi\,G_k^{(1)}\Big)^{-1}\,E_k\;\,\gb^{\m\n}\p_\m\,\Rb \, . 
\ee
We see that if $E_k\neq0$, the existence of a self-consistent background requires that the energy-momentum tensor is not conserved, $\Db_\m  T^{\m\n}\neq0$,
unless $\p_\m\,\Rb=0$. Using \eqref{4.35} we can re-express \eqref{4.39} as
\be\label{4.40}
\Db_\m  T^{\m\n}[A,\gb]=\frac{E_k}{2(1+2E_k)}\,\gb^{\m\n}\Db_\m\Big(\gb_{\a\b}  T^{\a\b}[A,\gb] \Big) \, .
\ee
Again, we see that $ T^{\m\n}$ can be conserved only if the trace $\gb_{\a\b}  T^{\a\b}$, and therefore $\Rb$, happens to be constant.

Equivalently, we could have started form the form \eqref{4.36} of the tadpole equation. Then the integrability condition is $\Db_\m \widetilde{T}^{\m\n}=0$
which, when rewritten in terms of $T^{\m\n}$, likewise leads to eq.\ \eqref{4.40}.

So, given this new kind of integrability condition at finite $k$, can we actually expect the tadpole equation to have solutions? First of all it is clear
that for $T^{\m\n}=0$ the effective Einstein equation is of the standard type, so for pure gravity\footnote{
While this case is not considered in the present paper, we expect that in pure quantum gravity, too, an $E_k$-term is generated.}
the situation is the same as in classical general relativity.

As for gravity coupled to matter, in classical relativity a standard argument shows that the energy-momentum tensor is conserved if the matter action is
invariant under diffeomorphisms {\it and the matter fields satisfy their equations of motion}. Remarkably, the same argument does not go through
for the tadpole equation at $k\neq0$, for two independent reasons. Indeed, as $\G_k^{\textrm{M}}[A,g,\gb]$ is diffeomorphism invariant it satisfies a Ward identity
of the form \eqref{1.8} which, with \eqref{4.34}, can be seen to imply at $g=\gb$
\be\label{4.41}
\Db_\m  T^{\m\n}[A,\gb]=-\frac{\gb^{\m\n}\p_\m A}{\Sgb}\,\frac{\d\G_k^{\textrm{M}}[A,\gb,\gb]}{\d A}+\frac{2}{\Sgb}\,\Db_\m\frac{\d\G_k^{\textrm{M}}[A,g,\gb]}{\d\gb_{\m\n} }\Bigg|_{g=\gb}.
\ee
Note that the first term on the RHS of \eqref{4.41} is not in general zero when $A$ is on shell, the reason being that \eqref{4.41} involves 
$\d\G_k^{\textrm{M}}/\d A$ while the equation of motion is \eqref{4.31}, or $\d\G_k^{\textrm{M}}/\d A=-\Sgb\,\cR_k[\gb]\,A$. Thus, this term equals 
$\gb^{\m\n}\,(\p_\m A)\,\cR_k[\gb]\,A$ and so it is nonzero even when $A$ is on-shell. However, as it should be, it vanishes in the limit $k\to0$.

The second term on the RHS of \eqref{4.41} is nonzero precisely when $\G_k^{\textrm{M}}$ has an ``extra'' $\gb$-dependence. As we discussed already, this dependence
must disappear for $k\to0$,  but at finite $k$ there is no reason why the RG evolution should not generate such a $\gb$-dependence.

So we can summarize the situation with matter by saying that, on  the one hand, the integrability of the tadpole equation generically requires a non-conserved
$T^{\m\n}$ at $k\neq0$, and that on the other hand the very structure of the average action entails  that the coarse graining and the RG running  actually do produce a
non-conservation of a certain type. Whether this generic non-conservation of $T^{\m\n}$ is such that it renders the tadpole equation integrable cannot
be assessed by general arguments but only in  special examples; we shall come back to this issue elsewhere.

It should also be emphasized that for the existence of an asymptotically safe fundamental theory it is by no means necessary that there are self-consistent backgrounds
for all physical situations. Such backgrounds are merely a convenient tool for the {\it visualization} of special properties and predictions of the theory, but
they are not needed for its {\it construction} in terms of an RG flow on theory space. It is quite conceivable that in a regime where the quantum effects are strong 
no self-consistent background will exist, indicating that the concept of a classical mean field metric breaks down there.

\section{Conclusion}
The gravitational average action approach to quantum gravity solves the ``background independence'' problem by giving an extra $\gb$-dependence to $\G_k$.
The average action $\G_k[g_{\m\n},\gb_{\m\n},\cdots]$ is inherently of a bimetric nature. In particular a functional flow equation which is exact in Wilson's sense
can be formulated only on a theory space of functionals depending on two metrics. Almost all investigations within this framework performed to date employ
 truncations which retain only the minimum extra  $\gb$-dependence due to the gauge fixing term. A comprehensive analysis of the extra background field dependence
is highly desirable, however, both for improving the quality and precision of the predictions, and for gaining insights into  various structural issues and problems
which are obscured in a single metric approximation. In particular in the Asymptotic Safety context bimetric truncations are likely to be unavoidable in order 
to improve upon the  precision that has been achieved already. In fact, in a first investigation based upon a bimetric ansatz \cite{elisa2} utilizing the conformally reduced gravity framework,
 it was found that the modifications caused by the generalization of the truncation are probably larger than the typical scheme
dependence within the original single metric truncation. Generalizing these investigations to full fledged gravity is difficult because of the complicated
functional traces appearing in the FRGE.

In the present paper we took the first step in this direction by computing the RG flow of a bimetric action $\G_k[A,g,\gb]$ for a set of 
scalar fields interacting with all irreducible components of the metric.
Yet, to simplify matters and to make the structural issues as transparent as possible, we neglected the contribution of the metric fluctuations to the corresponding 
beta functions. This system was known to display a non-Gaussian fixed point when analyzed within a single metric truncation \cite{perper1,robertoannph}. As for the Asymptotic Safety program,
 our most important finding is that the bimetric truncation, too, gives rise to such a fixed point, a very encouraging result indeed.

On the conceptual side, the discussions in this paper should have made it clear that in an RG context there is no such thing as ``the'' Newton's constant,
or ``the'' cosmological constant. At scales $k\neq 0 $ the split symmetry which ties together $\hb$ and $\gb$ is broken by the cutoff, and therefore an
expansion of $\G_k[A,\hb;\gb]$ in powers of $\hb$ contains an infinite set of different field monomials, and hence running couplings, which
degenerate to the same quantity at the physical point $k=0$. For instance, there is a whole zoo of cosmological constants $\L_k^{(0)},\L_k^{(1)},\L_k^{(2)},\cdots$.
In the  $\hb$-expansion of $\G_k$ they are the coefficients of the monomials which arise when we expand $\Sg\equiv \sqrt{\textrm{det}(\gb_{\m\n}+\hb_{\m\n})}$ with respect
to $\hb_{\m\n}$. Those monomials are characterized not only by the number of $\hb$'s they contain, but also by their tensor structure. At order $\hb^2$, for instance,
the two basis monomials are $(\gb^{\m\n}\hb_{\m\n})^2$ and  $\hb_{\m\n}\gb^{\m\r}\gb^{\n\s}\hb_{\r\s}$.

Similar remarks apply to the sectors with a nonzero number of derivatives acting on $\hb$ and/or $\gb$. If $I[g]=I[\gb+\hb]$ is any invariant built from a single
metric we can expand it in $\hb$ and arrive at a representation $I[\gb+\hb]=\sum_{n=0}^{\infty}u_{\textrm{class}}^{(n)}\;I^{(n)}[\hb;\gb]$.
Again, the expansion generates infinitely many diffeomorphically invariant monomials in $\hb$, $I^{(n)}[\hb;\gb]$. We assume that they are normalized in some canonical way
(``unit prefactor''). Hence the Taylor series gives rise to a well defined set of coefficients $u^{(n)}_{\textrm{class}}$ multiplying them.
It is convenient to take  the $I^{(n)}$'s as a subset of the basis elements spanning theory space. Then a single invariant $I[g]$, depending on one
metric, is seen to supply infinitely many terms to be included in $\G_k$, and these terms depend on $\hb$ and $\gb$ independently. As split symmetry is broken in general they
evolve differently under the RG flow.
Hence the corresponding part of $\G_k$ reads $\sum_{n=0}^{\infty}u_k^{(n)}I^{(n)}[\hb;\gb]$ with $k$-dependent coefficients $u_k^{(n)}$.
Only if split symmetry happens to be intact we have $u_k^{(n)}=u^{(n)}_{\textrm{class}}$ for all $n$. 

Imposing the restoration of split symmetry at $k=0$ provides a relation between these coupling constants, fixing the initial conditions $u_0^{(n)} = u_{\rm class}^{(n)}$ for all $n$, while coupling constants not associated to an expansion of $I[\gb+\hb]$ have to vanish. This reduces the undetermined couplings governing ``asymptotically safe'' RG
trajectories to the ones present in the single metric case. In other words, moving from the single metric to the bimetric setup does not result in a proliferation of coupling constants, as one could have expected initially.

In Sections 3 and \ref{sect.4} we made this structure explicit, working at the lowest nontrivial order of both the $\hb$- and the derivative-expansion.
We derived the RG flow of the terms linear in $\hb$, investigated its fixed point properties and, as an application, set up the tadpole equation
for self-consistent background geometries. Even at the two-derivative level this $k$-dependent field equation for $\gb$ does not exactly
have the structure of the classical Einstein equation. As  it cannot be obtained as the variation of any single metric action,
further tensor structures are possible. As demonstrated in Section \ref{sect.4}, the two-derivative approximation of the tadpole equation can be rewritten
in the style of Einstein's equation, but with a non-standard source term on its RHS. It also contains still another
 running cosmological constant, $\widetilde{\L}_k^{(1)}$.

In Section 2 we performed a complementary analysis, for constant metrics $g_{\m\n}$ and $\gb_{\m\n}$ only, but without assuming that $\hb=g-\gb$ is small.
Studying the RG flow of all possible non-derivative terms we saw that, contrary to common belief, it is not predominantly a $\Sg$ term which is
induced by the vacuum fluctuations of the matter fields, but rather $\Sgb$ and more complicated invariants involving both metrics. Even though
this observation by itself is not a solution to the cosmological constant problem it suggests that the possibility of vacuum fluctuations curving spacetime 
should be investigated with much more care.

\section*{Acknowledgements}
%
We thank G.\ 't Hooft and J.\ Pawlowski for interesting discussions. The research of E.M.\ and F.S.\ is supported by the Deutsche Forschungsgemeinschaft (DFG)
within the Emmy-Noether program (Grant SA/1975 1-1).

\newpage
\appendix
\section{Beta functions for the derivative- and $\hb$-expansion}
In this appendix we derive the RG equations \eqref{4.6} and \eqref{4.7} resulting from the truncation ansatz \eqref{4.1}.
\subsection{The master equation}
Using eq.\ \eqref{4.1} as the LHS of the FRGE \eqref{3.20}, we obtain the following ``master equation''
\begin{align}\label{A.1}
\frac{1}{8\pi}\int\textrm{d}^dx\Sgb \Bigg\{ \p_t & \left(\frac{\L_k^{(0)}}{G_k^{(0)}}\right)
+\frac{1}{2}\;\p_t\left(\frac{\L_k^{(1)}}{G_k^{(1)}}\right)\gb^{\m\n}\hb_{\m\n}
-\frac{1}{2}\;\p_t\left(\frac{1}{G_k^{(0)}}\right)\Rb\nonumber\\
{} & +\frac{1}{2}\;\p_t\left(\frac{1}{G_k^{(1)}}\right)\Gbb^{\m\n}\hb_{\m\n}
 - \frac{1}{4}\;\p_t\left(\frac{E_k}{G_k^{(1)}}\right)\Rb\;\gb^{\m\n}\,\hb_{\m\n}\Bigg\}\nonumber\\
=\frac{1}{2}\,\Tr  & \left[ \left(  \G^{(2)}_k[A,g=\gb+\hb,\gb]  +\cR_k[\gb]\right)^{-1}\;\p_t\cR_k[\gb]\right] \, ,
\end{align}
from which all beta-functions for the running coupling constants can be read off, by matching the coefficients of the $\hb$ and derivative expansion
on both sides. Thus we must expand its RHS in $\hb_{\m\n}$ and retain all terms of order zero and one.
The field $\hb_{\m\n}$ appears only inside the Hessian
\be\label{A.2}
\G^{(2)}_k[A,g=\gb+\hb,\gb]=\G^{(2)}_k[A,\gb,\gb] +\delta\G^{(2)}_k[\gb]+ O(\hb_{\m\n}^2) \, .
\ee
Here, by definition, $\delta\G^{(2)}_k[\gb]$ stands for the term linear in  $\hb_{\m\n}$. Upon inserting \eqref{A.2} into the
functional trace of \eqref{A.1} and expanding the denominator we see that the RHS of the master equation
\eqref{A.1} is given by the sum of the following two traces, containing the order zero and order one contributions, respectively
\begin{align}
T_0= & \frac{1}{2}\,\Tr \left[  \left(  \G^{(2)}_k[A,\gb,\gb]  +\cR_k[\gb]\right)^{-1}\;\p_t\cR_k[\gb] \right] \, , \label{A.3}  \\
T_1= &-\frac{1}{2}\, \Tr \left[  \left(  \G^{(2)}_k[A, \gb,\gb]  +\cR_k[\gb]\right)^{-2}\;\p_t\cR_k[\gb] \; \delta\G^{(2)}_k[\gb] \right] \, . \label{A.4}  
\end{align}
In the sequel, we will evaluate these traces up to first order in the background curvature, employing early-time heat-kernel techniques.
\subsection{Matrix elements of $ \G^{(2)}_k$ and $\delta\G^{(2)}_k$}
%
We start by computing the position space matrix elements \eqref{3.21} for the functional \eqref{4.1}. For $\hb\neq0$, i.e., $g\neq\gb$
\begin{align}\label{A.5}
\left(  \G^{(2)}_k[A, g,\gb]\right)_{xy}= & -\frac{1}{2}\frac{1}{\sqrt{\gb(x)}\sqrt{\gb(y)}}\Bigg[\sqrt{g(x)}\,\square_x\;\delta(x-y)
+\sqrt{g(y)}\,\square_y\;\delta(x-y)\Bigg]\nonumber  \\
{} & +\frac{\sqrt{g(x)}}{\gb(x)}\;\cU_{k}{\!\!''}(A(x))\;\delta(x-y) \, . 
\end{align}
Note the occurrence of both $g$ and $\gb$ in \eqref{A.5}. As always, $\square \equiv g^{\mu\nu} D_\mu D_\nu$ stands for the Laplace-Beltrami operator built from $g_{\m\n}$.
Further expanding the matrix elements \eqref{A.5} up to first order in $\hb_{\m\n}$ we obtain the linear term, with the abbreviation
$\gamma\=\half\gb^{\m\n}\hb_{\m\n}$, 
\begin{align}\label{A.6}
\left( \delta \G^{(2)}_k\right)_{xy}= & -\frac{1}{2}\left[\frac{\gamma(x)}{\sqrt{\gb(y)}}\,\bBox_x\;\delta(x-y)+ 
\frac{\gamma(y)}{\sqrt{\gb(x)}}\,\bBox_y\;\delta(x-y)\right]\nonumber\\
{} & -\frac{1}{2}\left[ \frac{1}{\sqrt{\gb(y)}}\,\delta\square_x\;\delta(x-y)+ 
\frac{1}{\sqrt{\gb(x)}}\,\delta\square_y\;\delta(x-y)\right]\Bigg|_{g=\gb}\nonumber\\
{} & + \frac{\gamma(x)}{\sqrt{\gb(x)}}\;\cU_{k}{\!\!''}(A(x))\;\delta(x-y) \, . 
\end{align}
Formally this result is obtained by applying a variation $\delta g_{\m\n}=\hb_{\m\n}$ to \eqref{A.5} and setting $g=\gb$ afterwards. The variation of 
the Laplacian, $\delta\square\big|_{g=\gb}$, is left unevaluated for the time being. In fact we shall see that there is a very elegant way of calculating
this type of traces without relying on explicit expressions. 
\subsection{Associated differential operators and IR-cutoff}
Applying the rule described in the paragraph following eq.\ \eqref{3.21} we can associate a differential operator to the matrix elements $\big( \G^{(2)}_k\big)_{xy}$.
For \eqref{A.5} this yields
\be\label{A.7}
\G^{(2)}_k[A, g,\gb]=\frac{\sqrt{g(x)}}{\sqrt{\gb(x)}}\;\Big[\,-\square\,+\,\cU_{k}{\!\!''}(A)\, \Big] \, .
\ee
In particular, for $g=\gb$,
\be\label{A.8}
\G^{(2)}_k[A, \gb,\gb]=-\bBox\,+\,\cU_{k}{\!\!''}(A) \, .
\ee
Analogously, the differential operator resulting from \eqref{A.6} is
\be\label{A.11}
\delta\, \G^{(2)}_k = -\frac{1}{2}\;\Bigg[\gamma(x)\,\bBox\;+\; \bBox\,\gamma(x) + 
\delta\square+(\delta\square)^{\dag}\Bigg]+ \gamma(x)\,\cU_{k}{\!\!''}(A) \, .
\ee
Here the hermitian adjoint operator $(\delta\square)^{\dag}$ is defined with respect to the inner product given by 
$(\cA_1,\cA_2)\=\int\textrm{d}^dx\sqrt{\gb(x)}\cA_1(x)\cA_2(x)$.

At this point a word of caution might be appropriate. While the {\it matrix elements} of $\delta\, \G^{(2)}_k$ can be obtained by
applying a formal variation ``$\delta$'' to those of $\G^{(2)}_k$, the same is not true for the associated {\it differential operators}.
If we naively apply a ``$\delta$'' to eq.\ \eqref{A.7} the result is {\it not} the (correct) equation \eqref{A.11}; in place of the
hermitian combinations $\gamma\bBox+\bBox\gamma$ and  $\delta\square+ (\delta\square)^{\dag}$, respectively, we would obtain
the non-hermitian operators $2\gamma\bBox$ and $2\delta\square$.

Considering \eqref{A.8}, we can see that the correctly adjusted cutoff operator is given by
\be\label{A.9}
\cR_k[\gb]=k^2\;R^{(0)}\left(-\tfrac{\bBox}{k^2}\right)\=\cR_k(-\bBox) \, , 
\ee
since then
\be\label{A.10}
\G^{(2)}_k[A, \gb,\gb]+\cR_k[\gb]=-\bBox\,+\,\cR_k[\gb]\,+\,\cU_{k}{\!\!''}(A) \, ,
\ee
so that the inverse propagator of the low momentum modes becomes $-\bBox+k^2+\cU_{k}{\!\!''}$, as it should be.

\subsection{The relevant traces}

Let $V:\Rom \to \Rom$ be an arbitrary real valued function. Then $V(D^2(g))^{\dag}=V(D^2(g))$ is hermitian with respect to $(\cdot\, , \cdot)$,
whence $\Tr [V(D^2(g))]$ is real for any metric $g_{\m\n}$. As a consequence, $\delta\Tr [V(D^2(g))]=\Tr[V'(D^2(\gb))\delta D^2]\=\Tr[V(D^2(\gb+\hb))]
-\Tr[V(D^2(\gb))]$ is real, too. Furthermore, since $\Tr[\cA^{\dag}]=\Tr[\cA]^*$ for any matrix $\cA$, we may write 
\begin{align}
\Tr\Big[V'(D^2(\gb))\delta D^2\Big] &=  \Tr\Big[\Big\{V'(D^2(\gb))\delta D^2\Big\}^{\dag}\Big]\nonumber\\
{} & =\Tr\Big[\big\{\delta D^2\big\}^{\dag}\big\{V'(D^2(\gb))\big\}^{\dag}\Big]\nonumber\\
{} & =\Tr\Big[V'(D^2(\gb))\big\{\delta D^2\big\}^{\dag}\Big]\nonumber \, .
\end{align}
This relation implies, for an arbitrary function $F\=V'$,
\be\label{A.12}
\frac{1}{2}\Tr\Big[F(\bBox)\Big(\delta\square+(\delta\square)^{\dag}\Big)\Big]=\Tr\Big[F(\bBox)\;\delta\square\Big] \, . 
\ee
Likewise exploiting the cyclicity of the trace we have
\be\label{A.13}
\frac{1}{2}\Tr\Big[F(\bBox)\Big(\gamma\bBox+\bBox\gamma\Big)\Big]=\Tr\Big[F(\bBox)\;\bBox\;\gamma\Big] \, . 
\ee

The identities \eqref{A.12} and \eqref{A.13} imply that under the trace \eqref{A.4} we may effectively replace the operator
$\delta\G^{(2)}$ of eq.\ \eqref{A.11} with $\big[-\bBox+\cU_{k}{\!\!''}(A) \big]\gamma- \delta\square$. As a consequence of this,
and also by using \eqref{A.11} we may  rewrite $T_0$ and $T_1\=T_1^A+T_1^B$ according to
\begin{align}
T_0= &  \frac{1}{2}\,\Tr\Bigg[ \Big(-\bBox + \cR_k(-\bBox)+ \cU_{k}{\!\!''}(A)\Big)^{-1}\p_t \cR_k(-\bBox)\Bigg]\label{A.14} \, , \\
T_1^A= &  -\frac{1}{4}\,\Tr\Bigg[ \Big(-\bBox + \cR_k(-\bBox)+ \cU_{k}{\!\!''}(A)\Big)^{-2}\p_t \cR_k(-\bBox)\Big(-\bBox
+ \cU_{k}{\!\!''}(A) \Big)\gb^{\m\n}\hb_{\m\n}\Bigg]\label{A.15} \, , \\
T_1^B= &  -\frac{1}{2}\,\Tr\Bigg[ \Big(-\bBox + \cR_k(-\bBox)+ \cU_{k}{\!\!''}(A)\Big)^{-2}\p_t \cR_k(-\bBox) \;\delta(-\square)\Bigg] \, . \label{A.16}
\end{align}
Thus the RHS of the master formula is given by the sum $T_0+T_1^A+T_1^B$.

\subsection{Explicit evaluation of the operator traces}
We are now in the position to compute the traces $T_0$, $T_1^A$, and $T_1^B$ explicitly. From now on we set $\cU_{k}{\!\!''}=\mb^2=$const, as in the main text.

\noindent
{\bf (A) Evaluation of $T_0$}\\
The trace $T_0$ of \eqref{A.14} provides the order zero contribution that occurred already in the single metric truncation.
 It is straightforwardly evaluated using the familiar heat kernel based expansion \cite{mr}
\be\label{A.17}
\Tr\big[W(-D^2)\big]= (4\pi)^{-d/2}\,\tr(I)\;\Bigg\{  Q_{d/2}[W]\int\textrm{d}^dx\Sg\,+\,\tfrac{1}{6}\,Q_{d/2-1}[W]\int\textrm{d}^dx\Sg\, R\,+\, O(R^2)        \Bigg\}
\ee
with $Q_n[W]\=\tfrac{1}{\G(n)}\int_0^{\io}\textrm{d}z\, z^{n-1}\, W(z)$ and $\tr(I)$ denoting the unit-trace with respect to the internal indices which, in our case, just counts the number of scalars. The result is most conveniently expressed in terms of the standard 
threshold functions \cite{mr}
\be\label{A.18}
\Phi_n^p(w)\=\frac{1}{\G(n)}\int_0^{\infty}\textrm{d}z\, z^{n-1}\;\frac{R^{(0)}(z)\,-\,z\,{R^{(0)}}'(z)}{\big[z\,+\,R^{(0)}(z)\,+\,w\big]^p} \, . 
\ee
Applying \eqref{A.17} to \eqref{A.14} yields, retaining terms with at most two derivatives of $\gb_{\m\n}$,
\be\label{A.19}
T_0=\frac{n_s}{(4\pi)^{d/2}}\Bigg\{ k^d\, \Phi^1_{d/2}(m^2)\int\textrm{d}^dx\Sgb\,+\,\tfrac{1}{6}\;k^{d-2}\;\Phi^1_{d/2-1}(m^2) \int\textrm{d}^dx\Sgb\, \Rb \Bigg\} \, . 
\ee

\noindent
{\bf (B) Evaluation of $T_1^A$}\\
The trace $T_1^A$ is also fairly simple to compute since the local form of \eqref{A.17} suffices for this purpose. In fact, when we evaluate the trace \eqref{A.15}
in the position space representation we may pull out the position dependent operator $\gb^{\m\n}(\hat{x})\hb_{\m\n}(\hat{x})$ from the matrix element by virtue of
the eigenvalue equation of $\hat{x}$:
\begin{align}\label{A.20}
T_1^A & = \int\textrm{d}^dx\;\langle x| \;\{\cdots \}\;\gb^{\m\n}(\hat{x})\hb_{\m\n}(\hat{x})\;|x\rangle\nonumber\\
{} & = \int\textrm{d}^dx\;\langle x|\; \{\cdots \}\;|x\rangle\;\gb^{\m\n}({x})\hb_{\m\n}({x}) \, . 
\end{align}
Using the local analog of equation  \eqref{A.17} for the diagonal matrix element $\langle x| \{\cdots \}|x\rangle$ we arrive at
\begin{align}\label{A.21}
T_1^A = & \, -\frac{n_s}{4(4\pi)^{d/2}}\,k^d\,\Big[ d\;\Phi^2_{d/2+1}(m^2)\,+\, 2m^2\;\Phi^2_{d/2}(m^2)\Big] \int\textrm{d}^dx\Sgb\,\gb^{\m\n}\hb_{\m\n}
\nonumber\\
{} & \,  -\frac{n_s}{24(4\pi)^{d/2}}\,k^{d-2}\,\Big[ (d-2)\;\Phi^2_{d/2}(m^2)\,+\, 2m^2\;\Phi^2_{d/2-1}(m^2)\Big] \int\textrm{d}^dx\Sgb\,\Rb\,\gb^{\m\n}\hb_{\m\n} \, . 
\end{align}

\noindent
{\bf (C) Evaluation of $T_1^B$} \\
In general the evaluation of traces involving factors of $\delta\square$ requires very lengthy computations and this is in fact the reason why we restrict
ourselves to the first order in $\hb_{\m\n}$ in the present paper. The trace \eqref{A.16} has the structure
$\Tr [\{\cdots\}\delta\square]$ with a single $\delta\square$ insertion only, and such traces can be calculated very elegantly by means of the following trick.
The expansion \eqref{A.17} is identically satisfied for all metrics $g_{\m\n}$. We may therefore perform the variation $g_{\m\n}\to g_{\m\n}+\delta g_{\m\n}$
on both sides of this expansion. After a short calculation the result is found to be
\begin{multline}\label{A.22}
\Tr\Big[W(-D^2)\delta (-D^2)  \Big]= \frac{1}{(4\pi)^{d/2}}\,\tr (I)\int\textrm{d}^dx\Sg\Bigg\{  -\half Q_{d/2+1}[W]\,g^{\m\n}\\
+\,\tfrac{1}{6}\,Q_{d/2}[W]\,\Big( G^{\m\n}\,-\,D^\m D^\n\,+\, g^{\m\n}D^2 \Big) \Bigg\}\delta  g_{\m\n} \, .
\end{multline}

With this identity in our hands it is now straightforward to calculate $T_1^B$. Applying \eqref{A.22} with $g_{\m\n}=\gb_{\m\n}$ and $\delta g_{\m\n}=\hb_{\m\n}$
to \eqref{A.16} we find, discarding surface terms,
\be\label{A.23}
\begin{split}
T_1^B=& \, \frac{n_s}{2(4\pi)^{d/2}}\,k^d\,\Phi^2_{d/2+1}(m^2) \int\textrm{d}^dx\Sgb\,\gb^{\m\n}\hb_{\m\n} \\
& \, -\frac{n_s}{6(4\pi)^{d/2}}\,k^{d-2}\,\Phi^2_{d/2}(m^2) \int\textrm{d}^dx\Sgb\,\Gbb^{\m\n}\hb_{\m\n} \, . 
\end{split}
\ee
The analogous computations at order $\hb_{\m\n}^2$  and higher are considerably more involved \cite{alcod}.

\noindent
{\bf (D) The beta functions} \\
The RHS of eq. \eqref{A.1} is explicitly known at this point; it is given by the sum of \eqref{A.19}, \eqref{A.21} and \eqref{A.23}.
By equating the coefficients of like field monomials on both sides of \eqref{A.1} we can now read off the beta functions for the
various combinations of coupling constants. Expressing the result in terms of dimensionless variables we thus arrive at the final
results given in eqs. \eqref{4.6} and   \eqref{4.7} of the main text.

\end{document}